\documentclass[12pt,a4paper,final]{iopart}
\usepackage{iopams}  
\usepackage{graphicx}
\usepackage[breaklinks=true,colorlinks=true,linkcolor=blue,urlcolor=blue,citecolor=blue]{hyperref}

\begin{document}
	
	\title[Decreasing groundwater quality at Cisadane riverbanks]
	{Decreasing groundwater quality at Cisadane riverbanks: groundwater-surface water approach}
	
	\author{Irawan, DE.$^1$, Puradimaja, DJ.$^1$, Yeni, D.$^3$, Kuntoro, AA.$^2$, and Julian, MM.$^1$}
	
	\address{$ˆ1$ Faculty of Earth Sciences and Technology, Institut Teknologi Bandung, 
		Jalan Ganesa No. 10, Bandung-40132, West Java, Indonesia}
	\address{$ˆ2$ Faculty of Civil and Environmental Engineering, Institut Teknologi Bandung, 
		Jalan Ganesa No. 10, Bandung-40132, West Java, Indonesia}
	\address{$ˆ3$ Agency for Environment of Tangerang Regency, Institut Teknologi Bandung, 
		Jalan Ganesa No. 10, Bandung-40132, West Java, Indonesia}
	\ead{erwin@fitb.itb.ac.id}
	
	%\maketitle
	
	\begin{abstract}
		
		The decreasing of groundwater quality has been the major issue in Tangerang area. One of the key process is the interaction between groundwater and Cisadane river water, which flows over volcanic deposits of Bojongmanik Fm, Genteng Fm, Tuf Banten, and Alluvial Fan. The objective of this study is to unravel such interactions based on the potentiometric mapping in the riverbank. We had 60 stop sites along the riverbank for groundwater and river water level observations, and chemical measurements (TDS, EC, temp, and pH). Three river water gauge were also analyzed to see the fluctuations.
		
		We identified three types of hydrodynamic relationships with fairly low flow gradients: effluent flow at Segmen I (Kranggan - Batuceper) with 0.2-0.25 gradient, perched flow at Segmen II (Batuceper-Kalibaru) with gradient 0.2-0.25, and influent flow at Segmen III (Kalibaru-Tanjungburung) with gradient 0.15-0.20. Such low flow gradient is controlled by the moderate to low morphological slope in the area. The gaining and losing stream model were also supported by the river water fluctuation data. TDS and EC readings increased more than 40$\%$ from upstream to downstream. At some points the both measurements were two times higher than the permissible limits, along with the drops of pH values at those areas.
		
		This study shows the very close interaction between Cisadane river water and groundwater in the riverbank. Therefore the authorities need to be managed the areas with a very strict regulations related to the small and large scale industries located near by the river.
		
	\end{abstract}
	
	\vspace{2pc}
	
	\maketitle
	
	\section{Introduction}
	
\begin{figure}[ht]
		\begin{center}
			\includegraphics[width=10cm]{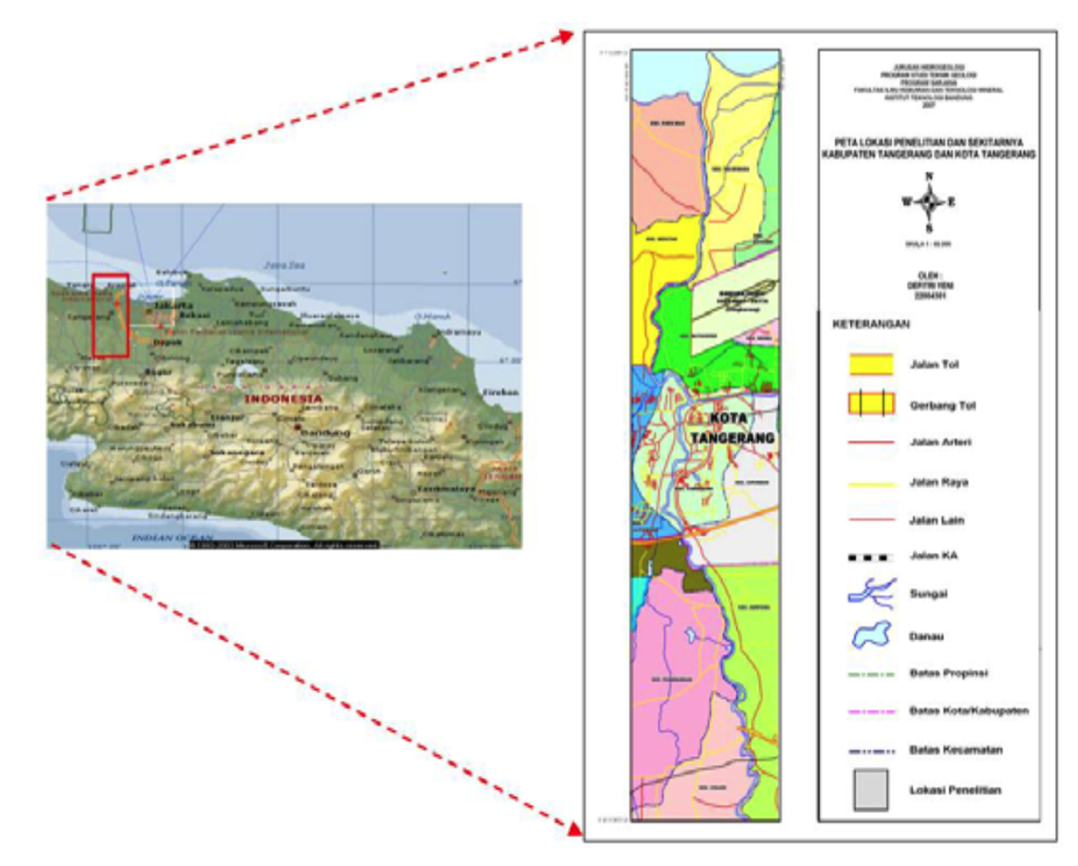}
		\end{center}
		\caption{Map of the study location}
		\label{location}
	\end{figure}

\begin{figure}[ht]
		\centering
		\includegraphics[width=10cm]{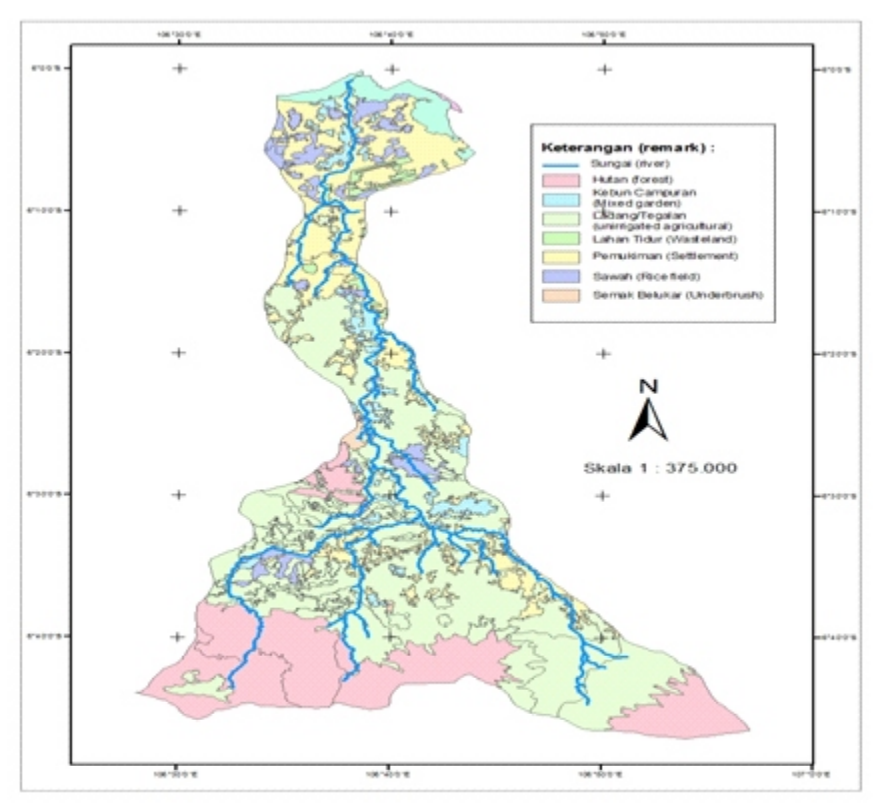}
		\caption{Profile showing the deteoriating water quality towards downstream}
		\label{watershed}
	\end{figure}

	In this area, groundwater has been used for domestic, industrial, agricultural and aquaculture purposes for more than 100 years. It has been highly exploited since 1990s due to the raise of urbanization and economy. The river flows through areas that has been developed as central of economy activities. It has to be supported by all means of water sources, including Cisadane (Figure \ref{location}) and Figure \ref{watershed}. 
	
	Major impact due to such condition is the decreasing of groundwater quality has been the major issue in Tangerang area as shown by organic and non-organic indicators\cite{Sukartaatmadja_2006,susana2010tingkat,muchtar2007nutrient,susilowati2007struktur,zarkasyi2008biosorpsi,junaidi2011pengaruh}. One of the key process is the interaction between groundwater and Cisadane river, as also shown by another case in Cikapundung river basin. The objective of this study is to unravel such decreasing as a function of water interactions between river water and groundwater in the riverbank. This information will be important for optimal use and sustainable management of the water resources in the area.

	\section{Materials and methods}
	
	The field investigation and sampling of river, canal and groundwater were performed in the dry season of 2006. Totally, 30 water samples were stored in 100 ml plastic bottles, consist of: 10 river water samples and 20 groundwater samples from shallow private tube-wells, with  well depth ranges from 10m to 30m below land surface. The sample dates are rather old but the results are still relevant with current situation. We propose newer research in 2011 by a team from Sam Ratulangi University \cite{siahaan_kualitas_2011} for comparison. 
	
	We conducted field measurements consist of groundwater levels, temperature, pH, Electrical Conductivity (EC), Total dissolved solids (TDS) using portable tools, Hanna Instruments. The samples collected were filtered and sent to the laboratory to be analysed to determine their Dissolved Oxygen (DO), major cations (sodium, potassium, calcium, magnesium) and major anions (chloride, sulfate, nitrate) by using inductively coupled plasma optical emission spectrometer (ICP-OES) and chromatography(IC). We also used Biological Oxygen Demand (BOD) and Chemical Oxygen Demand (COD) to obtain the organic setting in the environment.
	
\begin{figure}[ht]
		\centering
		\includegraphics[width=10cm]{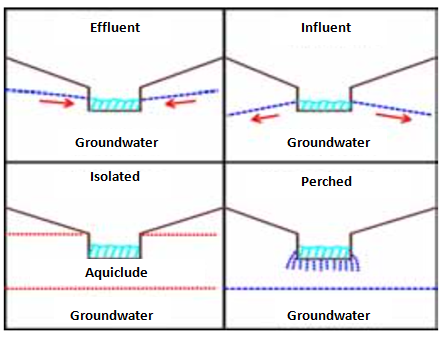}
		\caption{Profile showing the deteriorating water quality towards downstream}
		\label{model}
	\end{figure}

	Groundwater and river water level data were used to make potentiometric map. We plotted the data set and build a water flow model based on the contours using using GIS software. Subsequently, we intrepreted the interaction between groundwater and river water based on the following model (Figure \ref{model}), gathered from following  sources \cite{lubis_relasi_1997,zhou_groundwater-surface_2014,irawan_groundwatersurface_2014}.
	
	\section{Results and discussions}
	
\begin{figure}[ht]
		\centering
		\includegraphics[width=10cm]{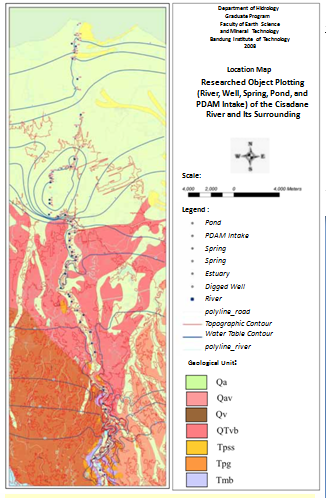}
		\caption{Profile showing the deteoriating water quality towards downstream}
		\label{flownet}
	\end{figure}

	Potentiometric mapping has shown three major river water-groundwater interactions (Figure \ref{flownet}) at three segments as follows: 
	
	\begin{enumerate}
		\item Effluent interaction (Segment 1, Kranggan – Batu Ceper) with characteristics:
		\begin{itemize}
			\item groundwater recharges to river,
			\item river water level elevation 10 -18.75 masl,
			\item groundwater level elevation 12.5- 30 masl,
			\item hydraulic gradient from 0.2 to 0.25.
		\end{itemize}
		\item Perched interaction (Segment 2, Batuceper-Kalibaru) with characteristics:
		\begin{itemize}
			\item river water infiltrates to aquifer in river bed. The distance between river bed and groundwater level ranges 1.25 – 7.5 m,
			\item river water level elevation 6.25 – 12.5 masl,
			\item groundwater level elevation 1 - 4 masl,
			\item hydraulic gradient from 0.2 to 0.25.
		\end{itemize}
		\item Influent interaction (Segment 3, Kalibaru-Tanjungburung) with characteristics:
		\begin{itemize}
			\item river water infiltrates to aquifer,
			\item river water level elevation 0 - 5 masl,
			\item groundwater level elevation 0 – 2 masl,
			\item hydraulic gradient from 0,15 to 0,2.
		\end{itemize}
	\end{enumerate}
	
\begin{figure}[ht]
		\centering
		\includegraphics[width=10cm]{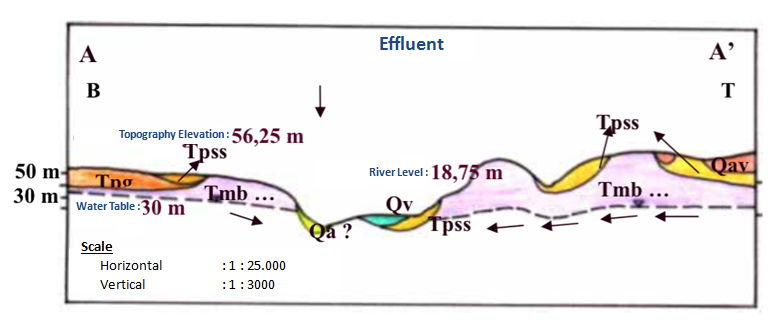}
		\includegraphics[width=10cm]{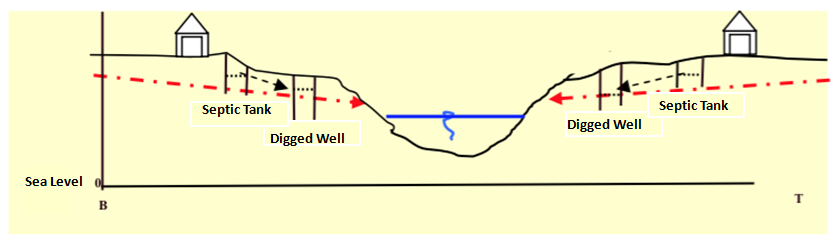}
		\caption{Example of subsurface profile in section 3}
		\label{section3}
	\end{figure}

	At all three segments, groundwater flow is controlled by relatively flat ground level with slope gradient between 0.016 to 0.02). Following such condition, groundwater flow gradient is also very low, from 0.15 to 0.25 (see Figure \ref{flownet}). Two geological sections have been built to give a short illustration on the subsurface condition of the area (Figure \ref{section3}). We can see variation of geometry of young alluvium deposit in all segments with fairly random in thickness. 
	
\begin{figure}[ht]
		\centering
		\includegraphics[width=10cm]{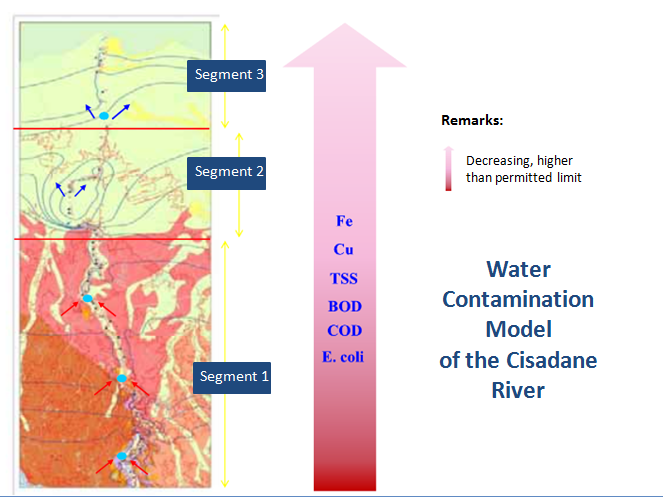}
		\caption{Profile showing the deteoriating water quality towards downstream}
		\label{qualityprofile}
	\end{figure}

	The deterioration of river and groundwater quality is shown by the increasing concentrations of Iron, Copper, TSS, BOD, COD, and E. coli towards downstream (Figure \ref{qualityprofile}). Such  contamination is due to the increasing settlement and industrial activities along the river. Related publications \cite{siahaan_kualitas_2011,rochyatun_distribusi_2006,wijaya_komunitas_2009} confirm such condition. Average TSS content 73.38 mg/l (higher than max limit of 50 mg/l) portrays high natural erosion, and also man-made sand digging sites in the Kranggan area. Solids from aerated Fe$^{2+}$ and Cu$^{+}$ can also contribute to the TSS value. The samples also show averagely high iron content 0.61 mg/l (max limit: 0.30 mg/l) and Copper 0.13 mg/l (max limit: 0.02 mg/l) which possibly came from many electroplating industries in the riverbank, as one of the source. Organic condition is indicated by average BOD value 8.42 mg/l (limit: 2 mg/l), COD 25.75 mg/l (limit 10 mg/l), and E. coli 6275/100 liter (limit 1000/100 liter). All three indicators are higher than tolerable limits, which came from poor sanitation and domestic waste management. Similar situation is shown in Cikapundung river \cite{Darul_2015,Darul_2016,Sancayaningsih_2015,Surtikanti_2005,muchtar2007nutrient}.
	
	\section{Conclusions}
	
	The results from this paper can be used to explain water quality deterioration along Cisadane riverbank. Three unique groundwater and river water interactions have been detected using relatively simple and cheap method. Major element concentrations were able to show the quality transition from upstream to downstream. However more samples should be added with more heavy or trace minerals measurements such as As and Pb for more detail water quality zones and involving processes. 
	
	The above-mentioned situation has been un-treated. Many efforts in years have been done to overcome the contamination, but the impact is still not significant. Our results in 2006 is still comparable with what another team had identified in 2011. Therefore we need to strongly address the condition to be continuously treated by the local government and inter-governments collaboration.     	
	
	\section*{Acknowledgement}
	
	This research was supported by Bandung Institue of Technology Research Grant. We are grateful to Office for Environment Tangerang Regency for their assistance and our field team from Department of Geology for their continuous support. We also thankful to Dr. Ratna Siahaan from Sam Ratulangi University, as one of the previous researcher in the area, for her review on the first draft.
	
	\section*{References}
	
	\bibliographystyle{plain}
	\bibliography{cisadane.bib}

\begin{thebibliography}{10}

\bibitem{Darul_2015}
A.~Darul, D.~E. Irawan, and N.~J. Trilaksono.
\newblock Groundwater and river water interaction on cikapundung river:
  Revisited.
\newblock {AIP} Publishing, 2015.

\bibitem{Darul_2016}
Achmad Darul, Dasapta~Erwin Irawan, Nurjanna~Joko Trilaksono, Aditya Pratama,
  and Ulfi~Rizki Fitria.
\newblock Conceptual model of groundwater and river water interactions in
  cikapundung riverbank, bandung, west java.
\newblock {\em {IOP} Conf. Ser.: Earth Environ. Sci.}, 29:012026, jan 2016.

\bibitem{irawan_groundwatersurface_2014}
D.E. Irawan, H.~Silaen, P.~Sumintadireja, R.F. Lubis, B.~Brahmantyo, and D.J.
  Puradimaja.
\newblock Groundwater–surface water interactions of {Ciliwung} {River}
  streams, segment {Bogor}–{Jakarta}, {Indonesia}.
\newblock {\em Environmental Earth Sciences}, pages 1--8, 2014.

\bibitem{junaidi2011pengaruh}
Edy Junaidi and Surya~Dharma Tarigan.
\newblock Pengaruh hutan dalam pengaturan tata air dan proses sedimentasi
  daerah aliran sungai (das): Studi kasus di das cisadane.
\newblock {\em Junal Rehabilitasi Hutan Dan Konservasi Alam}, 8(2):155--176,
  2011.

\bibitem{lubis_relasi_1997}
Rachmat~F Lubis.
\newblock {\em Relasi hidrodinamika air tanah dan air sungai di {Bantaran}
  {Sungai} {Cikapundung}, {Segmen} {Maribaya}-{Dayeuhkolot}}.
\newblock master thesis, Department of Geology, Institut Teknologi Bandung,
  Bandung, 1997.

\bibitem{muchtar2007nutrient}
Muswerry Muchtar.
\newblock Nutrient concentration and ph in the cisadane estuarine and coastal
  waters.
\newblock {\em MRI Marine Research in Indonesia}, page~63, 2007.

\bibitem{rochyatun_distribusi_2006}
Endang Rochyatun, M.~Taufik Kaisupy, and Abdul Rozak.
\newblock Distribusi logam berat dalam air dan sedimen di perairan muara sungai
  {Cisadane}.
\newblock {\em Makara Sains}, 10(1):35--40, 2006.

\bibitem{Sancayaningsih_2015}
Retno~Peni Sancayaningsih and Amarizni Mosyaftiani.
\newblock Vegetaion analysis in part of catchment area influencing water
  quality in cikapundung upstream, suntenjaya village, west bandung regency.
\newblock {\em {KnE} Life Sciences}, 2(1):234, sep 2015.

\bibitem{siahaan_kualitas_2011}
Ratna Siahaan, Andry Indrawan, Dedi Soedharma, and Lilik~B. Prasetyo.
\newblock Kualitas air s. cisadane jawa barat-banten.
\newblock {\em Jurnal Ilmiah Sains}, 11(2):268--273, 2011.

\bibitem{Sukartaatmadja_2006}
Sukandi Sukartaatmadja.
\newblock Evaluasi aliran permukaan, erosi dan sedimentasi di sub das cisadane
  hulu dengan menggunakan model {AGNPS} (agricultural non point souce pollution
  model).
\newblock {\em {JTEP}}, 20(3):217--223, dec 2006.

\bibitem{Surtikanti_2005}
Hertien Surtikanti.
\newblock Kesehatan lingkungan di das cikapundung akibat pencemaran air.
\newblock {\em Jurnal Pengajaran Matematika dan Ilmu Pengetahuan Alam}, 6(2):9,
  dec 2005.

\bibitem{susana2010tingkat}
Tjutju Susana.
\newblock Tingkat keasaman (ph) dan oksigen terlarut sebagai indikator kualitas
  perairan sekitar muara sungai cisadane.
\newblock {\em Jurnal Teknologi Lingkungan Universitas Trisakti}, 5(2):pp--33,
  2010.

\bibitem{susilowati2007struktur}
Emi Susilowati.
\newblock Struktur komunikasi makrozoobenthos sebagai indikator biologi
  perairan di hulu sungai cisadane, bogor.
\newblock {\em IPB repository}, 2007.

\bibitem{wijaya_komunitas_2009}
Habib~Krisna Wijaya.
\newblock Komunitas {Perifiton} dan {Fitoplankton} serta {Parameter}
  {Fisiska}-{Kimia} {Perairan} sebagai {Penentu} {Kualitas} {Air} di {Bagian}
  {Hulu} {Sungai} {Cisadane}, {Jawa} {Barat}.
\newblock {\em IPB repository}, 2009.

\bibitem{zarkasyi2008biosorpsi}
Hafidh Zarkasyi.
\newblock Biosorpsi logam merkuri (hg) oleh bacillus megaterium asal hilir
  sungai cisadane.
\newblock {\em UNJ repository}, 2008.

\bibitem{zhou_groundwater-surface_2014}
S.a Zhou, X.a~b Yuan, S.a Peng, J.a Yue, X.a Wang, H.a Liu, and D.D.c Williams.
\newblock Groundwater-surface water interactions in the hyporheic zone under
  climate change scenarios.
\newblock {\em Environmental Science and Pollution Research}, 2014.
\newblock cited By (since 1996)0; Article in Press.

\end{thebibliography}

\end{document}